\documentclass[twocolumn]{aastex62}
\usepackage{subcaption}
\usepackage{amsmath}
\usepackage[utf8]{inputenc}
\usepackage{xcolor}
\usepackage{amssymb}
\usepackage{graphicx}
\usepackage{url}

\begin{document}
\newcommand{\cowork}{\it \color{blue}}
\newcommand{\be}{\begin{equation}}
\newcommand{\ee}{\end{equation}}
\newcommand{\bea}{\begin{equation}\begin{aligned}}
\newcommand{\eea}{\end{aligned}\end{equation}}
\def\msol{M_\odot}\def\msol{M_\odot}
\newcommand{\reviewer}{}

\title{Flares from merged magnetars: \\their prospects as a new population of gamma-ray counterparts of binary neutron star mergers}
\author{Shu-Xu Yi}
\thanks{sxyi@ihep.ac.cn}
\author{Zhen Zhang}
\thanks{zhangzhen@ihep.ac.cn}
\author{Xilu Wang}
\thanks{wangxl@ihep.ac.cn}
\affil{Key Laboratory of Particle Astrophysics, Institute of High Energy Physics, Chinese Academy of Sciences,\ 19B Yuquan Road, Beijing 100049, China}
\date{October 2022}
\begin{abstract}
   Long-lived massive magnetars are expected to be remnants of some binary neutron star (BNS) mergers. In this paper, we argue that the magnetic powered flaring activities of these merged magnetars would occur dominantly in their early millisecond-period-spin phase, which is in the timescale of days. Such flares endure significant absorption by the ejecta from the BNS collision, and their detectable energy range is from 0.1-10 MeV, in a time-lag of $\sim$ days after the merger events indicated by the gravitational wave chirps. We estimate the rate of such flares in different energy ranges, and find that there could have been ~0.1-10 cases detected by Fermi/GBM. A careful search for $\sim10$ milliseconds spin period modulation in weak short gamma-ray bursts (GRBs) may identify them from the archival data. The next generation MeV detectors could detect them at a mildly higher rate. The recent report on the Quasi-Period-Oscillation found in two BASTE GRBs should not be considered as cases of such flares, for they were detected in a lower energy range and with a much shorter period spin modulation. 
   \end{abstract}
\section{Introduction}
Magnetars are a kind of neutron stars (NSs) which have extremely strong magnetic fields. The magnetar's magnetic field can be as strong as $\sim10^{13-15}$ G \citep{2008MNRAS.389L..66F,2011ASSP...21..247R,2015SSRv..191..315M,2015RPPh...78k6901T,2017ARA&A..55..261K}, while that of an ordinary NS is $\sim10^{10-12}$ G (but see ``low-magnetic field" magnetars \citep{2010Sci...330..944R,2013IJMPD..2230024T}). The typical radiation activities are believed to be powered by the huge energy reservoir in the magnetic fields of magnetars, rather than their rotational energy or gravitational energy as those in spin powered or accretion powered NSs. Such magnetar radiation activities were observed as anomalous X-ray pulsars (AXPs) and Soft Gamma-ray Repeaters (SGRs). AXPs appear to be isolated pulsars with X-ray emission, whose spin down luminosity are thought to be insufficient to power their observed luminosity \citep{1981Natur.293..202F,2002Natur.419..142G,2004NuPhS.132..456K}; SGRs are thought to be magnetars which give off bursts in gamma-ray at irregular time intervals \citep{1984Natur.307...41G,1991ApJ...366..240N,1995A&ARv...6..225H,1995MNRAS.275..255T,2008A&ARv..15..225M}. Besides, there are rare cases, where much more energetic flares are emitted from magnetars, which are referred to as ``Giant Flares" \citep{2005Natur.434.1107P,2005Natur.434.1098H,2020AstL...46..573M,2020ApJ...903L..32Z,2021AAS...23723302R,2021Natur.589..211S}. 

The latter two flaring activities are believed to originate from the release of magnetic energy during occasional magnetic field reconnection in magnetars. There are various theories to explain the underlying triggers of such recombination. Following the dichotomy of \cite{2023arXivS}, the first class of mechanisms attributes the trigger to crustal destructive or defective events (the ``star quake" paradigm, see models e.g., \citealt{1989ApJ...343..839B,1996ApJ...473..322T,2006MNRAS.368L..35L,2020ApJ...896..142B,2020ApJ...897..173B,2020ApJ...900L..21Y}); while the second attributes to the reconfiguration in the twisted magnetosphere due to magnetohydrodynamical (MHD) instabilities\footnote{Specific MHD instabilities to be expected in this scenario are tearing mode instability \citep{2003MNRAS.346..540L,2007MNRAS.374..415K}, plasmoid instability \citep{2019MNRAS.485..299R}, sausage/varicose mode and kink instability \citep{2011ApJ...735L..20L} \textit{etc.}.} (the ``solar flare" paradigm, e.g., \citealt{2003MNRAS.346..540L,2007MNRAS.374..415K,2019MNRAS.485..299R,2019MNRAS.490.4858M}).

Unlike those isolated evolved magnetars, there is a population of magnetars that were born in the remnants of binary neutron star (BNS) collisions. We refer to these magnetars as merged-magnetars, which are the focus of this manuscript. It is widely believed that, BNS collision, if the remnant is not massive enough to cause a prompt collapse into a black hole (BH), will result a massive magnetar in millisecond time scale \citep{ 1992ApJ...392L...9D,1992Natur.357..472U,1994MNRAS.270..480T,1998ApJ...494L.163Y,1998ApJ...498L..31B,1998ApJ...505L.113K,1998PThPh.100..921N,1999A&A...341L...1S,2000ApJ...537..810W,2000ApJ...542..243R,2014PhRvD..90d1502K}. Such a magnetar inherits most of the orbital angular momentum of the progenitor NSs, therefore initially possessing a short spin period of milliseconds, much shorter than the typical seconds-long spin period of magnetars. Because of their faster spin, larger mass and younger age, it is intuitive to suspect that they are even less stable than isolated evolved magnetars, and therefore, could be more likely to emit gamma-ray flares. 

In this work, we investigate the possibility that flares from merged-magnetars be identified, especially as electromagnetic wave counterparts (EMC) of gravitational wave (GW) signals of BNS mergers. The manuscript is arranged as follows: In section \ref{sec:II}, we give semi-quantitative arguments that the flare activities should dominantly occur in the merged magnetar's early phase. Then in the next section, the detectable event rate of merged-magnetar flares is estimated from a population of BNS mergers. In section \ref{sec:IV}, we consider their potential to be identified as EMC of GW, followed by a section where the absorption from ejected matter is taken into consideration. We discuss several relevant aspects and summarize the main findings in the last two sections. 

\section{Flaring activities in early and later phases of massive magnetars}
\label{sec:II}
 
\begin{figure}[ht]
\begin{subfigure}{0.4\textwidth}
  \centering
  \includegraphics[width=.8\linewidth]{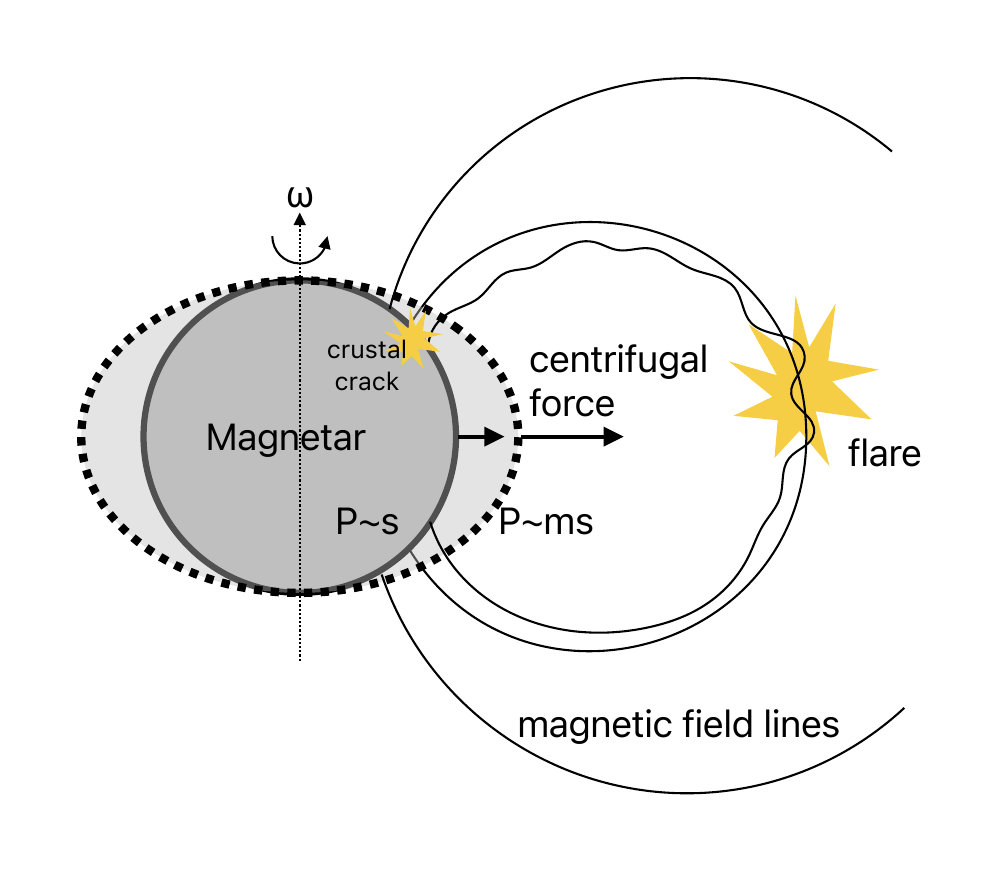}  
  \caption{Crust crack scenarios: Crustal defective events occurs more frequently in the stage when the magnetar fast spins down, since the centrifugal force which is counteracting the NS' self-gravity is rapidly decreasing. }
  \label{fig:sub-first}
\end{subfigure}
\begin{subfigure}{0.46\textwidth}
  \centering
  \includegraphics[width=.8\linewidth]{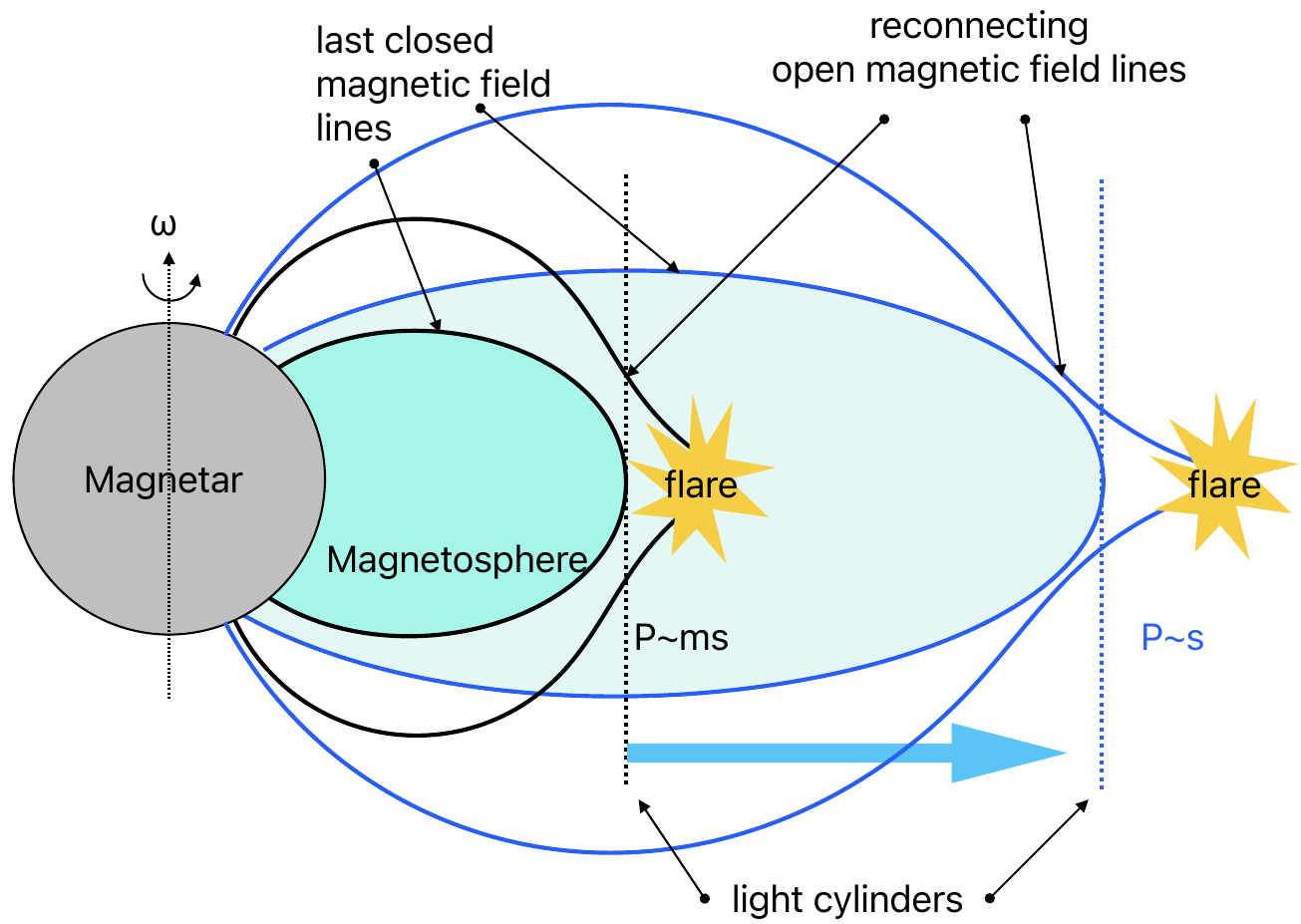}  
  \caption{Magnetosphere instability scenarios: Instabilities will be triggered frequently in the stage when the magnetar fast spins down, the boundary of the magnetar's magnetosphere is rapidly expanding.}
  \label{fig:sub-second}
\end{subfigure}
\caption{Illustrations of two scenarios in which the fast spinning-down millisecond magnetar is more probable to have gamma-ray flares}
\label{fig:fig}
\end{figure}

In the ``star quake" paradigm, where gamma-ray flares are triggered by crustal defective events, when the magnetar is fast spinning down due to the gigantic magnetic breaking torque, the centrifugal force decreases rapidly; see Figure~\ref{fig:fig} (a). The crust of the NS will re-adjust to its new equilibrium configuration, where the centrifugal force is counteracting against the self-gravity. In this re-configuring process, crustal defective events can be expect to be much frequent than when the NS enters the later slowly spinning-down stage (such spin-down induced star quake was first discuss by \citealt{1971AnPhy..66..816B}); in the second paradigm, where gamma-ray flares are attributed to the instabilities in the magnetosphere, the boundary of the magnetar's magnetosphere (near the light cylinder, whose radius $R_{\rm LC}=cP$, where $c$ is the velocity of light, and $P$ is the spin period) expand fast as its spin period rapidly increase; see Figure~\ref{fig:fig} (b). The magnetic field lines near the boundary will have to re-adjust accordingly, and thus are expected to be more likely to give-off gamma-ray flares\footnote{unlike in those conventional ``solar flare" paradigm, where the magnetic twist leaks from the NS interior into the magnetosphere, through the slow crustal deformation, here the magnetic twist fast expands along with the magnetosphere. Therefore we refer to it as the ``magnetosphere instability" paradigm for distinction}. In fact, from our following semi-quantitative estimation we show that, the rate of gamma-ray flares in the fast spinning-down millisecond period phase is much higher than its second period phase, so that the total flare energy released in the former phase (with the time scale of days) is comparable or higher than the latter phase (with the time scale of 10$^5$ years).

In the crust crack scenarios, let us assume that each crustal destructive event which is energetic enough to trigger a magnetar gamma-ray flare has a characteristic energy $E_{\rm crack}$. Denote the change in the centrifugal force as $\Delta F_{\rm cent}$ in a small interval of time $\Delta t$. The corresponding linear deformation of the NS is $\Delta l$, which we suppose is proportional to $\Delta F_{\rm cent}$ as $$\Delta l=\frac{\Delta F_{\rm cent}}{\kappa},$$ where $\kappa$ is the elastic factor of the crust. The work done by the gravity is thus:
\begin{equation}
    \Delta E=\frac{\Delta F_{\rm cent}}{\kappa}F_{\rm G},
\end{equation}
where $F_{\rm G}$ denote the gravitational force acting on the crust, which remains almost constant as the deformation is small. If we equalize the work done by gravity with that released in crust cracks, we have the following equation:
\begin{equation}
    \frac{\Delta F_{\rm cent}}{\kappa}F_{\rm G}=\Delta N_{\rm crack}E_{\rm crack},
    \label{eq:crackenergy}
\end{equation}
where $\Delta N_{\rm crack}$ is the number of crust cracks in $\Delta t$, where their ratio is the magnetar gamma-ray flare rate $$R_{\rm B}=\frac{\Delta N_{\rm crack}}{\Delta t},$$ which, according to equation \eqref{eq:crackenergy}, has the following proportional relationship:
\begin{equation}
    R_{\rm B}\propto\dot{F}_{\rm cent}\propto\frac{\dot{P}}{P^3}.
    \label{eq:ratecrack}
\end{equation}
Note that if the spinning-down torque is dominated by magnetic dipole braking\footnote{GW braking will only dominate over the magnetic braking when the spin period is less than 0.1 s \citep{1992Natur.357..472U,1998ApJ...498L..31B}. In this case, GW braking will fast spin-down the magnetar to a regime where the magnetic braking takes over \citep{2001ApJ...552L..35Z}.}, then we have:
\begin{equation}
    B^2_{\rm s}\propto P\dot{P},
\end{equation}
where $B_{\rm s}$ the surface magnetic field strength, we assume to be an constant. As a result, equation \eqref{eq:ratecrack} becomes:
\begin{equation}
    R_{\rm B}\propto\frac{1}{P^4}.
\end{equation}
Now, since in the first phase, the spin period is in the order of $\sim10$ ms, and in the second phase it is $\sim$s, the $R_{\rm B}$ in the first phase can be eight orders of magnitude larger than that in the second phase. On the other hand, the time-span of the first phase is about $10^{-8}-10^{-7}$ of that of the second phase. Therefore, the magnetar gamma-ray flare energy releasing in the first phase is the same or one order of magnitude larger than that in the second phase, as we claimed in above. 

If we are in the magnetosphere instability scenarios, in a short time interval $\Delta t$, the boundary of the magnetosphere (where close and open magnetic field lines transit) expand in distance: $\Delta R_{\rm LC}$. The volume which has been swept is:
\begin{equation}
    \Delta V=4\pi R_{\rm LC}^2\Delta R_{\rm LC}.
\end{equation}
The volume times the magnetic field energy density is the energy got involved. This energy is likely to be released by processes such as magnetic field re-connection near the light cylinder. We have:
\begin{equation}
    \Delta E\propto B_{r=R_{\rm LC}}^2R_{\rm LC}^2\Delta R_{\rm LC},
    \label{eq:energy_mag}
\end{equation}
where $B_{r=R_{\rm LC}}$ is the magnetic field strength at the light cylinder. For a dipole magnetic field, $$B_{r=R_{\rm LC}}\propto\frac{B_{\rm s}~}{R_{\rm LC}^3}.$$ Consequently, equation \eqref{eq:energy_mag} can be reformed into:
\begin{equation}
    \Delta E\propto B^2_{\rm s}\frac{\Delta R_{\rm LC}}{R^4_{\rm LC}}\propto \frac{\dot{P}}{P^4}\propto\frac{1}{P^5}.
\end{equation}
Therefore, the number of bursts with characteristic energy $E_{\rm{burst}}$ within $\Delta t$, \textit{i.e.}, the characteristic burst rate $R_{\rm{B}}$ is:
\begin{equation}
    R_{\rm{B}}=\frac{\Delta N_{\rm{burst}}}{\Delta t}=\frac{\Delta E/E_{\rm{burst}}}{\Delta t}\propto\frac{1}{P^5}.
\end{equation}
Using the similar argument as in the crust crack scenarios, we can see the ratio of the $R_{\rm B}$ between the first and second phase can be ten orders of magnitude, and thus the ratio of the corresponding total energy releasing can be $10^3$ in the magnetosphere instability scenarios. 

\section{The rate of flares from the population of merged magnetars}
\label{sec:III}

Define the magnetar flare energy differential number density distribution from a single merged-magnetar as: $n(E)$, where $E$ is the energy release during the flare. The total number of flares above some certain energy limit ($E_{\rm limit}$) during the life time of the magnetar is:
\begin{equation}
    N_{\rm B}=\int_{E_{\rm limit}}^\infty n(E)dE.
\end{equation}
and the total energy released is:
\begin{equation}
    E_{\rm B}=\int_0^\infty n(E)EdE.
\end{equation}
which should be less than the total energy stored in the magnetosphere. 

Now the rate of bursts from all merged-magnetars in the local Universe within a sphere shell of radius from $D$ to $D+dD$, in the energy range from $E$ to $E+dE$ is: 
\begin{equation}
    d^2R_{B}=4\pi D^2dDR_{\rm m}n(E)dE.
    \label{eq:RBE}
\end{equation}
where $R_{\rm m}$ is the merger rate density of double neutron stars {\it whose remnants are NSs instead of prompt collapsed BHs.} 
The above equation can be further formulated to:
\begin{equation}
    d^2R_{B}=4\pi D^2dDR_{\rm m}n(E)\frac{dE}{dF}dF,
    \label{eq:RF}
\end{equation}
where $F$ is the fluence. Since $E=4\pi D^2F$, we have from the above equation that:
\begin{equation}
    \frac{d^2R_{\rm B}}{dFdD}=(4\pi D^2)^2dDR_{\rm m}n(E).
\end{equation}
As a result, the rate of such bursts with in a limiting distance $D_{\rm u}$ and above a limiting fluence is:
\begin{equation}
    R_{\rm B}=
    (4\pi)^2R_{\rm m}\int_{F_{\rm limit}}^\infty\int_0^{D_{\rm u}}n(E(F))D^4dDdF
    \label{eq:keyeq}
\end{equation}
where $F_{\rm limit}$ is the fluence limit of the gamma-ray detector. 

The volumetric integral in equation \eqref{eq:RBE} should be limited to in local Universe, where the merger rate density can be viewed as a constant, and cosmic expansion has a negligible effect. When considering the joint observation of such flares with GW detection of the BNS mergers, the integral over the luminosity distance in equation \eqref{eq:RF} should be truncated at the BNS horizon of the GW detector.

The key quantities are {$n(\tau, E)$} and $R_{\rm m}$, where we reformulate the latter: $$R_{\rm m}=\eta\mathcal{R}_{\rm m}$$, where $\mathcal{R}_{\rm m}$ is the merger rate density of all BNS population, and $\eta$ is the fraction of those have long-lived magnetar remnants. $\mathcal{R}_{\rm m}$ can be constrained by previous GW observation at $39-1900$ Gpc$^{-3}$s$^{-1}$ \citep{2021arXiv211103634T}. 

We assume a power-law form of $n(E)$:
\begin{equation}
    n(E)=N_{\rm{B}}f_0E^{-\beta}, E_l<E<E_u. 
\end{equation}
Studies \citep{2020MNRAS.491.1498C} found the index in broad consistency with that expected from a Self-Organized Criticality (SOC, see \textit{e.g.,} \citealt{1987PhRvL..59..381B,1991ApJ...380L..89L,1992PhRvL..68.1244O,2011soca.book.....A}) process ($\beta=5/3$), and the factor $$f_0=(\beta-1)E_l^{\beta-1}$$

Now the equation {\eqref{eq:keyeq}} can be simplified to:
\begin{equation}
    R_{\rm B}=R_{\rm m}(4\pi)^{2-\beta}N_{\rm B}f_0\frac{F^{1-\beta}_{\rm limit}}{(\beta-1)}\frac{D^{5-2\beta}_{\rm u}}{(5-2\beta)}
    \label{eq:simcore}
\end{equation}

The total energy to be released should be limited by the magnetic energy stored in the magnetosphere:
\begin{equation}
    E_{\rm mag}\geq\int_{E_l}^{E_u}En(E)dE\sim N_{\rm B}\overline{E},
    \label{eq:Emag}
\end{equation}
\begin{equation}
    \overline{E}=\int_{E_l}^{E_u}\frac{n(E)}{N_{\rm{B}}}EdE\sim\frac{\beta-1}{2-\beta}\big(\frac{E_u}{E_l}\big)^{1-\beta}E_u.
    \label{eq:Ebar}
\end{equation}
the approximant in the above equation is valid only when $1<\beta<2$. \cite{2020MNRAS.491.1498C} found $\beta\sim1.66$, which meets the above mentioned conditions. If we assume that the flares in ordinary SGR and GFs follows the same energy distribution law, the energy of those flares can range more than five orders of magnitudes, with the $E_{\rm u}$ corresponding to the most energetic giant flare being $E\sim10^{46}$\,ergs. Therefore, from equation~\eqref{eq:Ebar} we find that: $\overline{E}\sim4\times10^{43}E^{\beta-1}_{l, 42}\,\rm ergs$, where $E_{l, 42}$ is the lower energy end of the $n(E)$ in unit of $10^{42}$\,ergs. 

The magnetic energy stored in the magnetosphere is \citep{ZYetal2022}:
\begin{equation}
    E_{\rm mag}\sim8\times10^{46}B^2_{15}\,\rm ergs,
\end{equation}
where $B$ is the surface magnetic field strength of the magnetar scaled with $10^{15}$ G. 
If we equalize the both sides of inequality \eqref{eq:Emag}, we can have a rough estimation of $N_{\rm B}$ as: 
\begin{equation}
    N_{\rm B}\sim2\times10^3\frac{B^2_{15}\,\rm{ergs}}{E^{\beta-1}_{l,42}},
\end{equation}
Taking the $N_{\rm B}$ from above, and taking $E_l=10^{42}$ergs, with the expression of $f_0$ into equation \eqref{eq:simcore}, we have:
\begin{equation}
    R_{\rm B}=2\times10^3\frac{(4\pi)^{2-\beta}}{5-2\beta}R_{\rm m}D^3_{\rm u}\big(\frac{10^{41}\,\text{ergs}}{F_{\rm limit}D^2_{\rm u}}\big)^{\beta-1}.
\end{equation}
If we insert the numbers into the above equation with $\beta=5/3$, we obtain that:
\begin{equation}
    R_{\rm B}=5\times10^{-3}\eta\mathcal{R}_{\rm m}B^2_{15}D_{u,100}^{5/3}F^{-2/3}_{\rm{limit},-8}\,\text{yr}^{-1},
    \label{eq:therate}
\end{equation}
where $\mathcal{R}_{\rm m}$ is in unit of yr$^{-1}$/Gpc$^3$, $D_{u,100}$ is the distance limit in unit of 100 Mpc, $F_{\rm{limit}, -8}$ is the fluence limit in unit of $10^{-8}$\,ergs/cm$^2$. The detection horizon $D_{u}$ is limited by the fluence cut of a gamma-ray detector as:
\begin{equation}
    D_{\rm u, \gamma}\sim300\,F_{\rm limit,-8}^{-1/2}\,\text{Mpc}
    \label{eq:theDgamma}
\end{equation}
when taking the $E_{u}=10^{46}$\,ergs, which corresponds to a conservative estimation of the total magnetic energy stored in the magnetosphere \citep{ZYetal2022}. It should be noted that, since we equalized inequality \ref{eq:Emag}, the estimated occurrence rate $R_{\rm B}$ is an upper limit.

\section{Merged-magnetar flares as EM counterpart of GW events and its spin period modulation}
\label{sec:IV}


A magnetar which was born with a millisecond spin period will experience two evolutionary phases. In its first phase of millisecond period of spinning, the magnetar's spin period rapidly slows down to seconds by the strong magnetic braking torque. In the later phase, the spin period is settled to seconds scale and evolves less rapidly. We can define a transition time between the first and the second phases as:
\begin{equation}
    \tau_{\rm trans}\sim\,\frac{P_{10\,\rm{ms}}^2}{B^2_{0,15}}\,{\rm day}.
\end{equation}

As argued in previous sections, the burst rate before $\tau_{\rm trans}$ overwhelms that after it. Therefore, the rate in equation \eqref{eq:therate} mostly describes those bursts occuring before $\tau_{\rm trans}$. Equivalently, those bursts to be detected is likely to following a GW chirp from BNS merger with a time lag $\tau_{\rm lag}<\tau_{\rm trans}$. On the other hand, $\tau_{\rm lag}$ should be larger than $\tau_{\rm limit}$, which
is the time limit less than which, the ejecta from the BNS merger is still optically thick, thus the flares from the magnetars will be largely absorbed and the temporal structure within the flares is smeared. $\tau_{\rm limit}$ is also in time scale of days (\citealt{1998ApJ...507L..59L}, and see discussion in following section). 

Since the duration of a typical magnetar GF is $\sim0.1$-1\,s, the flares detected in this phase can exhibit significant spin modulation, which can serve as an unambiguous evidence of the existence of a merged magnetar. In this case, the $D_{\rm u}$ in equation~\eqref{eq:therate} is the minimum between the gamma-ray detection horizon and the GW horizon:
\begin{equation}
    D_{u}=\min\big(D_{\rm u,\gamma}, D_{\rm GW}\big)
\end{equation}

\begin{figure}[!h]
    \centering
    \includegraphics[width=0.5\textwidth]{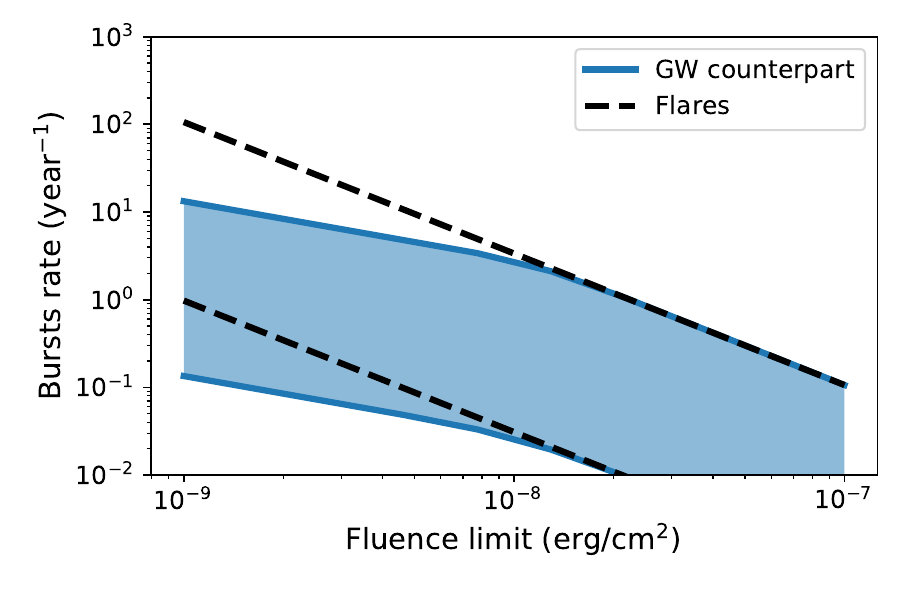}
    \caption{The flare rate as function of fluence limit is plotted in Figure \ref{fig:first}. Note that the occurrence rate is calculated under a strong assumption that the entire magnetic field energy in the magnetosphere is released as flares.} The blue band denotes the possible range of bursts rate which are associated with GW observation in LVK O4, and the dashed dark lines indicate those of bursts regardless of GW counterparts. The upper and lower limits of the range correspond to the 86\% quantiles (1-sigma) in a Morte Carlo simulation. 
    \label{fig:first}
\end{figure}

The flare rate as a function of fluence limit is plotted in Figure \ref{fig:first}, see figure caption for the detailed description of the plot. When plotting Figure \ref{fig:first}, we calculate the rate using equation~\eqref{eq:therate} with Monte Carlo samplings of $\eta$, $R_{\rm m}$ and $D_{\rm GW}$: $\eta$ is uniformly randomly sampled in log-space from 0.01 to 0.1; $R_{\rm}$ is sampled from a log-Gaussian distribution with 1-$\sigma$ upper and lower limits correspond to 39 and 1900\,yr$^{-1}$/Gpc$^3$; $D_{\rm GW}$ is sampled from a Gaussian random with mean 300\,Mpc and a standard deviation of 40\,Mpc, which corresponds to the BNS detection horizon of a GW detector network with LIGO-Virgo-KAGRA (LVK) in O4 period\footnote{as simulated here: \url{https://emfollow.docs.ligo.org/userguide/capabilities.html}}. 

\section{Absorption by the BNS ejecta}
\label{sec:V}

During the collision of BNS, abundant material will be ejected from both the tidal tail and the disk \citep{Bovard, Just}. Actually BNS are the confirmed site for rapid neutron capture nucleosynthesis  ($r$-process) \citep{Abbott, NSM, Cowperthwaite2017, Kasen2017}, which is responsible for about half of the elements heavier than iron measured in our solar system \citep{Burbidge1957, Sneden2008}.
Thus, it is expected that the BNS will be surrounded by dense $r$-process material at early time, with a total ejected mass ranging from $\sim 0.005-0.1 \msol$ \citep{Bovard, Radice18, CoteGW170817, Just, Rodrigo2015}, and is optically thick to the flare gamma-ray radiation from the center remnant due to the absorption/interaction processes of photons when propagating through the material, including Compton scattering, photoelectric absorption, pair production, \textit{etc}.. In the mean time, as the $r$-process material is ejected from BNS merger with high speed, the ejecta will become optically thin at $\sim$ days after the merging event \citep{1998ApJ...507L..59L, Oleg, Wang2020b}. The kilonova models of GW170817 observation suggested that such ejecta has a speed ranging from $0.1c$ to $0.3c$ on average \citep[e.g.,][]{Kasen2017, Rosswog2018, Wollaeger2018, Watson2019}, as expected from previous theoretical work \citep[e.g.,][]{1998ApJ...507L..59L, Tanaka2013}.

Our calculation in the previous section did not include the absorption of the surrounding BNS ejecta to the flare radiation. When such effect is included, together with a finite work energy range of the detector, it effectively replaces the limiting fluence in equations \eqref{eq:therate} and \eqref{eq:theDgamma} with a new limiting fluence $\Tilde{F}_{\rm limit}$, which is related with the original $F_{\rm limit}$ as:
\begin{equation}
    \Tilde{F}_{\rm limit}=\frac{F_{\rm limit}}{1-\xi}
\end{equation}
Here we define an absorption factor $\xi$ to describe the effect of the surrounding ejecta in absorbing the high-energy photons from the BNS flare, which is a function of time after BNS collision ($\tau$) and is defined as:
\begin{equation}
    1-\xi(\tau)=\frac{F_{\rm observed}(\tau)}{F_{\rm emitted}}=\frac{\int^{E_{\rm high}}_{E_{\rm low}}f_{\rm observed, E}(\tau)dE}{\int^{E_{\rm high}}_{E_{\rm low}}f_{E}dE},
    \label{eq:thexi}
\end{equation}
which is the ratio between the flux after absorption (observed) at time $\tau$ and the total emitted flux from the flare, and $E_{\rm low}$ and $E_{\rm high}$ denote the energy range where a specific detector works. $f_E=dE_{\gamma}/dEdAdt$ is the differential energy flux emitted from the flare.
We assume a spectrum shape of $f_E$ as a power law of index -0.2 with an exponentially-cutoff at $0.48$\,MeV, i.e., $f_E=f_0E^{-0.2}\exp(-E/0.48\rm{MeV})$, and $f_0$ is the normalization factor with $\int^{\infty}_{0}f_EdE=F_{\rm total}=L_{\rm total}/4\pi D^2$, where $D$ is the distance of the BNS, and $L_{\rm total}$ is the luminosity of the magnetar flare. This spectrum shape is taken from that of the GF from magnetar SGR 1806-20 \citep{2005Natur.434.1107P}.

For approximation, we assume a uniform spherical $r$-process ejecta distribution as in \citet{Wang2020a, Wang2020b} to calculate the observed flare emission (after the ejecta absorption). Then, the emitted gamma rays after propagation through the ejecta (due to various photon interactions in the ejecta) is 
\begin{equation}
f_{\rm observed, E}(\tau)=f_{E}e^{-\rho_{\rm ej}\kappa(E)l}
    \label{eq:oberved}
\end{equation}
where $\rho_{\rm ej}$ is the ejecta density, $\kappa(E)$ is the opacity of the ejecta to a photon with energy $E$, path-length $l$ is the distance of the photons travelling through the ejecta, in this case, $l\sim v\tau$, with $v$ to be the expanding/ejected velocity of the ejecta.
Only non-interacted photons are included in the observed gamma-ray signal here; scattered photons are ignored as their effects are minimal at late times when the ejecta is nearly optically thin.

\begin{figure}[htb!]
  \centering
  \includegraphics[width=1\linewidth]{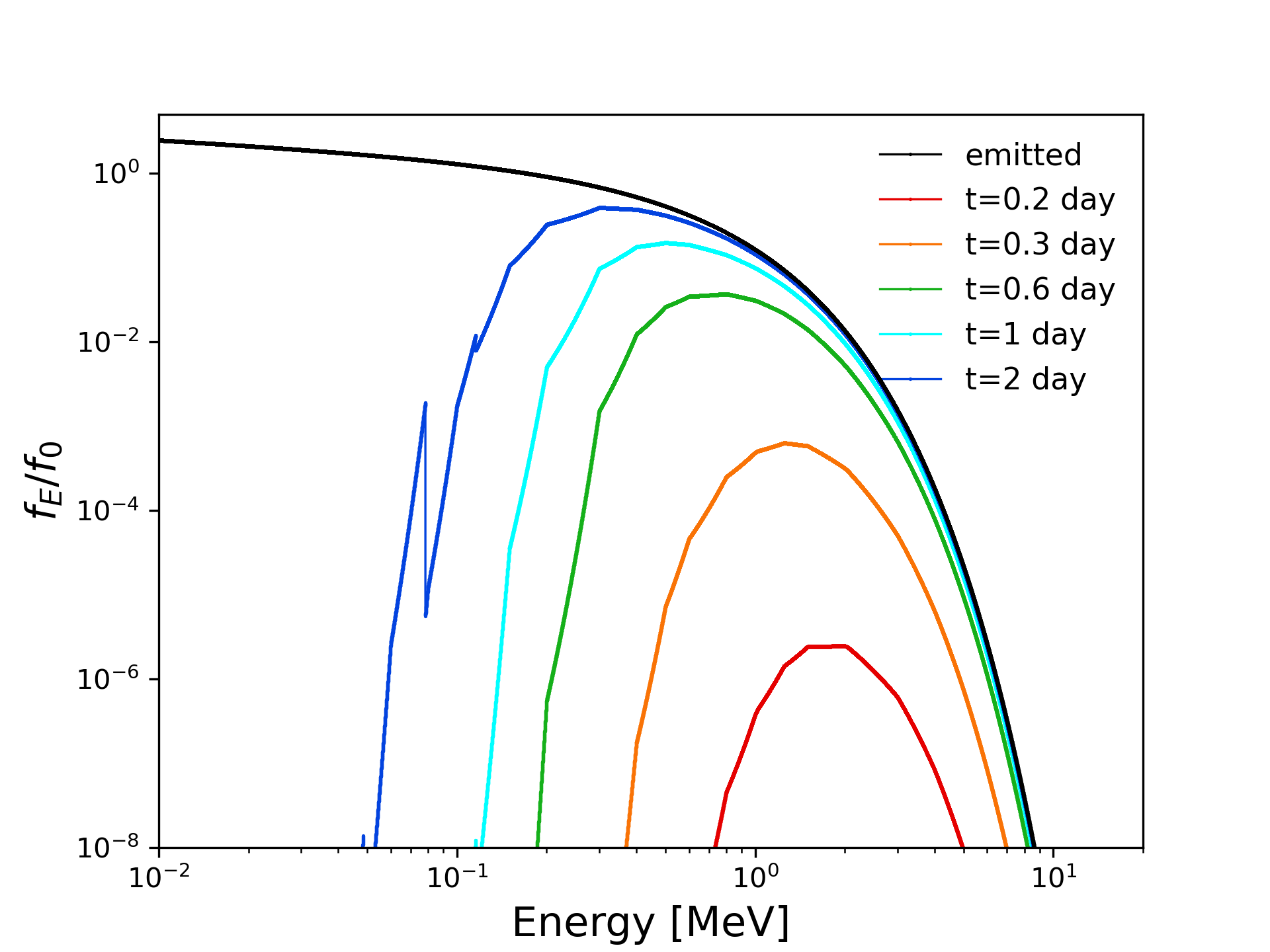} 
  \caption{Plot of the flare spectra flux $f_E/f_0$ vs energy E before (emitted, black) and after absorption by a 0.01 $\msol$ BNS merger ejecta with 0.3c expansion velocity and a robust main $r$-process components, at 5 different times after BNS collision: 0.2 day (red), 0.3 day (orange), 0.6 day (green), 1 day (cyan), 2 day (blue).}
  \label{fig:spectrum}
\end{figure}

We obtain the $r$-process nuclei abundance distribution in the BNS merger ejecta using the nuclear reaction network code Portable Routines for Integrated nucleoSynthesis Modeling, or PRISM \citep{Matt} as in \citet{Wang2020a,Wang2020b}. We adopt a BNS merger dynamical ejecta with robust $r$-process productions \citep{rosswog} for the baseline calculation. 
The opacity values for the total BNS collision ejecta are calculated using the ejecta's composition with a mixture of the opacity values of five characteristic isotopes (Fe, Xe, Eu, Pt, and U), the detailed procedure is described in \citet{Wang2020b}.
The opacity values of individual $r$-process nuclei are adopted from the XCOM website\footnote{\url{https://www.nist.gov/pml/xcom-photon-cross-sections-database}}, with photon interactions including coherent (Rayleigh) scattering, incoherent (Compton) scattering, photoelectric absorption, and pair production. For photons with energy above $\sim10$ MeV, the main interaction with the ejecta material is pair production; while at energy range $\sim0.5-\sim10$ MeV, the dominant process is incoherent (Compton) scattering; for lower energy photons (below $\sim0.5$ MeV), photoelectric process mostly takes place, manifested as edges shown at energy below $\sim0.5$ MeV in Figure~\ref{fig:spectrum}, especially for the blue line that reach the energy below 0.1 MeV. Such feature arises from {\reviewer spikes in} the BNS ejecta opacity/cross section at the same photon energies.
The resulting spectra after absorption by the ejecta at 5 different times (0.2 days, 0.3 days, 0.6 days, 1 day and 2 days) after BNS collision are shown in figure~\ref{fig:spectrum}. Compared to the emitted spectrum shown as a black line, we conclude that the detection window of such bursts should be in the energy range from $\sim$1\,MeV to 10\,MeV, and in the time window between 0.5 and 2 days after BNS collision.

We note that in addition to the burst flare radiation, the BNS collision ejecta itself also emit gamma-ray photons through the decays of the radioactive $r$-process nuclei.
The total gamma radiation rate from the $r$-process ejecta is estimated to be $\epsilon_0(\tau)\sim2\times 10^{10} {\rm erg\ g}^{-1} s^{-1} (\tau/{\rm day})^{-1/3}$ \citep{Metzger,Oleg}, and the $r$-process gamma-ray energy at $\sim1$day is then $\sim10^{41}$ erg/s for a 0.01 $\msol$ BNS merger ejecta. Thus, such signal would be small compared to the flare emissions, and the BNS ejecta spectrum shapes with nuclear decay lines \citep{Oleg, Wang2020b} are also different from the flare signal discussed here. 

\begin{figure}[htb!]
  \centering
  \includegraphics[width=1\linewidth]{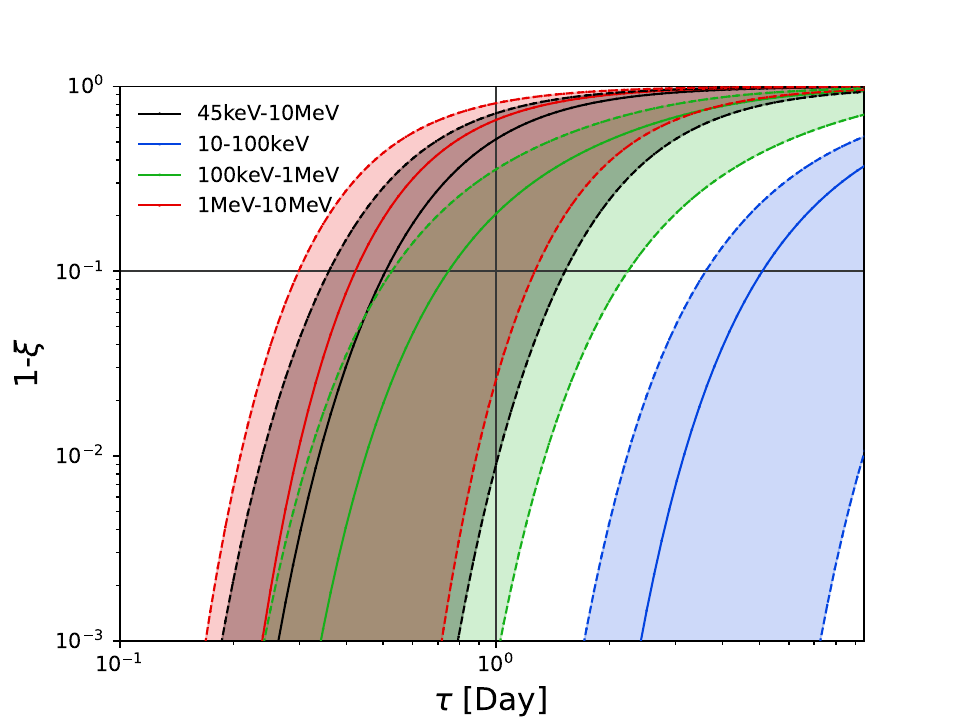} 
  \caption{Plot of the absorption factor $\xi$ versus time ($\tau$) for four different energy bands: 45keV-10MeV (black); 10-100keV(blue); 100keV-1MeV(green); 1MeV-10MeV(red). The solid lines are the absorption results for a 0.01 $\msol$ BNS ejecta with 0.3c expansion velocity and a robust main $r$-process components, the color shades indicate uncertainties due to the variations in the BNS ejecta properties including mass, velocity and composition, the dashed lines denote boundaries of $1-\xi$ for four different energy bands. {\reviewer We include two reference lines: $\tau=1 $\,day (vertical) and $1-\xi=0.1$ (horizontal). These lines represent the fiducial spin-down timescale and the point at which the ejecta transitions from being optically thick to optically thin, respectively.}}
  \label{fig:abs-factor}
\end{figure}

\begin{figure}
    \centering
    \includegraphics[width=1\linewidth]{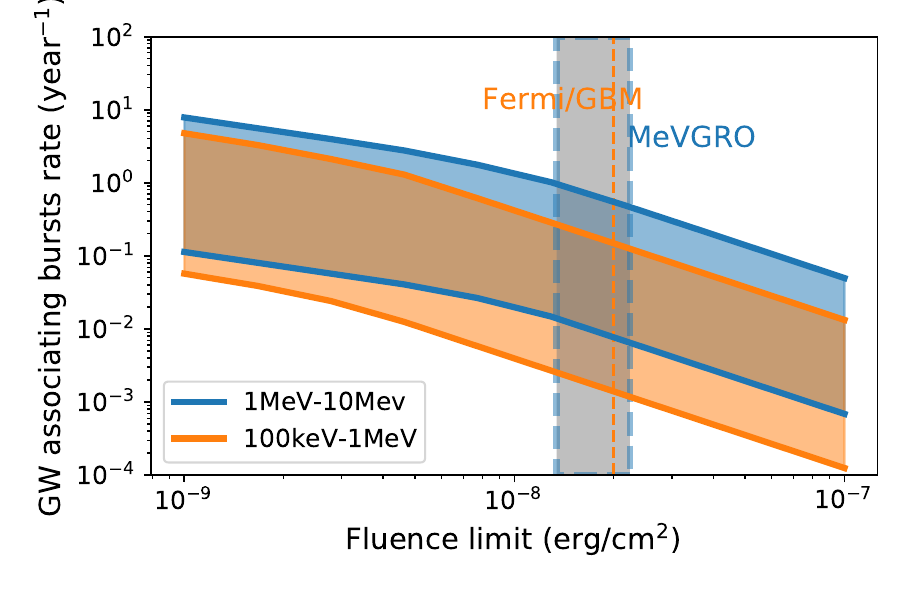}
    \caption{Same plot as the band for GW counterpart in Figure \ref{fig:first}, but with absorption considered for energy ranges from 1\,MeV to 10\,MeV and from 100\,keV to 1\,MeV. The orange dashed line is the estimated Fermi/GBM detector fluence limit for a 1s burst; while the vertical shaded region is the estimated fluence limit range for next generation MeV detectors like MeVGRO with $\sim$1s exposure time.}
    \label{fig:second}
\end{figure}

Then we conduct the integral in equation~\eqref{eq:thexi} to obtain $\xi$ as function of $\tau$ in Figure~\ref{fig:abs-factor}. The uncertainty bands are due to variations in BNS ejecta properties, including velocity, ejecta mass, and the components. Here we varied the ejecta mass between 0.005 to 0.03 $\msol$, and the velocity between 0.1c to 0.3c. To test the sensitivity of the signal to the ejecta component, we adopted the parameterized BNS outflow conditions \citep{Just, Radice18} with a range of initial electron fractions as in \citet{Wang2020b}, so that the ejecta component varies from the weak $r$-process (no third peak and heavier actinides elements) to robust $r$-process (with actinides). From Figure~\ref{fig:abs-factor}, we can see that at the detection window discussed above, the corresponding absorption factor is  $\xi\sim0.5-1$ with an order of magnitude uncertainty. The burst detection rate after absorption of BNS ejecta considered is re-plotted in Figure \ref{fig:second}. When plotting the figure, we calculate the rate with a $\xi$ randomly drawn from its corresponding range evaluated above with uniform distribution. 

During the variation test, we find that the flare signal is more sensitive to the ejecta velocity and mass. Therefore, on the other hand, the detection rate obtained in the real observation could enable us to put a constraint on the BNS ejecta property.



\section{Discussion}
\subsection{Potential cases in archival data from Fermi/GBM sGRB catalogue}
The Fermi/GBM detector has an energy range of $\sim0.01-1$\,MeV, and a fluence limit (for a 1\,s bursts) of $\sim2\times10^{-8}$\,erg/cm$^2$ \footnote{Following the practice of \cite{2022arXiv220814156H}, we use the lowest observed fluence of the sGRB catalogue \citep{2016ApJS..223...28N} as the fluence limit of second duration bursts.}. It has been monitoring GRBs for $\sim10$\,years. From our estimation, there should have been $\sim0.1-10$ such bursts detected in its bursts catalogue. As mentioned above, such flares may exhibit spin modulation. Although there has been searches for QPOs in song of Fermi/GBM's bright GRBs \citep{2013ApJ...777..132D} with no positive results, a more careful survey focusing on those weak short bursts with a fast increasing period $\sim0.1$\,s might identify such bursts in the archival data, although with foreseeable difficulties due to their fewer photon counts. Recently, \cite{2023Natur.613..253C} reported the detection of kHz QPOs in two archival sGRBs of the Burst and Transient Source Experiment (BATSE). BATSE works in lower energy range from 50 KeV to 300 keV, in which we expect significant absorption if they were merged-magnetar-flares. Besides, the found QPO are above 1\,kHz, which is much higher than we expect for the stable merged-magnetar-flares. Therefore, these two sGRBs with QPOs should not be considered as cases of the proposed merger-magnetar-flares. 
\subsection{Prospects for the next generation MeV detectors}
For the next generation MeV telescope, such as COSI\footnote{\url{https://cosi.ssl.berkeley.edu}}, AMEGO\footnote{\url{https://asd.gsfc.nasa.gov/amego/}}, or MeVGRO\footnote{\url{https://indico.icranet.org/event/1/contributions/777/}}, the energy range between $\sim0.1-10$ MeV will be well covered, and the detectors' sensitivities at this energy range are expected to be at least 1-2 orders of magnitude better than current and previous MeV detectors like INTEGRAL\footnote{\url{https://www.cosmos.esa.int/web/integral}} and COMPTEL\footnote{\url{https://heasarc.gsfc.nasa.gov/docs/cgro/comptel/}}. However, as those detectors are not specially designed to monitor burst sources, their sensitivities to second-duration transients are much {\reviewer fainter than} their reported value for continuous sources. Take MeVGRO as an example of the next generation MeV detectors, its fluence limit range estimated for $\sim1$ second observing time is represented in Figure \ref{fig:second} as blue bands. We can tell that, using the next generation MeV detectors such as MeVGRO, the detection rate of the merged magnetar flares can only be mildly larger, or similar to that using Fermi/GBM. 

The main sources of the uncertainties in the burst rate estimation is from: 1) the local rate density of the BNS collision; 2) the fraction of BNS merger which leaves long-lived magnetar, or $\xi$ as we denote in this paper; A future detection of a population of such bursts, together with multi-messenger observation from GW will in turn give inference on these aspects. The current BNS $\mathcal{R}_{\rm{m}}$ estimation is based on two BNS events (GW170817 and GW200311\_115853) in LIGO-Virgo-Collaboration (LVC) O2-O3 period \citep{2021arXiv211103634T}. In the LVK-O4 period, it is estimated that $36^{+49}_{-22}$ BNS mergers shall be detected\footnote{as reported in  \url{https://emfollow.docs.ligo.org/userguide/capabilities.html\#datadrivenexpectations}, using the same methodology in \cite{2020LRR....23....3A} but with updated input models for detector network and sources.}, which will return much tighter constraints on $\mathcal{R}_{\rm{m}}$. The value of $\xi$ depends on the equation of state of NS matter, and also the mass function of NS. The latter can be better constraints by a larger sample of BNS observed with GWs. If multiple bursts could be observed from a single merger magnetar, the dependence of the bursts properties on its period spin could be studied, which will provide valuable insights into the emission mechanism of magnetar activities. 
\subsection{Solidification of the crust of the newly born magnetar}
In order to justify the domination of the magnetar activity in its rapid braking stage in age of $\sim$days under the star quake paradigm, we made an order-of-magnitude estimation of the burst rate in the introduction section. A crucial presumption was that the elastic property of the neutron star crust remains unchanged. First we need to check whether the surface of the NS has already cooled enough to have a solidified crust. According to \cite{1973NuPhA.207..298N,1994A&A...283..313H,2000PhLB..485..107D}, the melting temperature of the NS crust lies well above 10$^8$\,K, and \cite{1994ApJ...425..802L} showed that the core of a newly born NS can fast cool down to temperature $T$ on the timescale: 
\begin{equation}
t=20\,(T/10^9\,\text{K})^{-4}\,\text{s}.
\end{equation}
Therefore, in the age of $\sim$\,day, the new born magnetar already has a solid crust. As for its elastic properties, their are in general not constants and temperature dependent (\textit{e.g.} \citealt{1991ApJ...375..679S}). Therefore, a more careful quantitative calculation of the burst rate as function of the magnetar's age should consider a realistic cooling curve of the NS's crust and its temperature-dependent elastic properties.  
\subsection{{\reviewer Distinguishing} the flaring paradigms with Galactic magnetars}
The semi-quantitative argument in section \ref{sec:II} leads to an interesting observation: different flaring paradigm predicts a distinct flaring rate as a function of the magnetar age. More specifically, in both paradigms the characteristic flaring rate is $R_{\rm{B}}\propto P^{-m}$, where $m=4,5$ for the crust crack and magnetosphere instability scenarios, respectively. Assuming that the surface magnetic field strength remains constant, and the spin down is solely contributed by magnetic dipole radiation braking, then $\dot{P}\propto 1/P$. Therefore, the characteristic age of {\reviewer a} magnetic is:
\begin{equation}
    \tau=\frac{P}{2\dot{P}}\propto P^2.
\end{equation}
As a result,
\begin{equation}
    R_{\rm{B}}\propto\tau^{-m/2}.
\end{equation}
One would {\reviewer therefore} expect that by comparing the observed characteristic burst rate of the flaring activity of Galactic magnetars against prediction, it could be possible to distinct flaring paradigms.

\section{Summary}
From our above argument and calculation, we conclude that flares from merged massive magnetars can be expected as a population of gamma-ray transients, which associate GW chirp events of BNS mergers. Such a gamma-ray counterpart of GW may look like a short gamma-ray bursts (sGRB) according to its duration, but it can be found with several distinct features: 
\begin{itemize}
    \item it tends to be weak in flux, and the time-lag between the burst and GW chirp is $\sim$1-2 days, rather than $\sim$\,s as in sGRB;
    \item its spectrum has a lower energy cut at $\sim$100\,keV, due to absorption of the ejecta.
\end{itemize}
Besides, it may show spin modulation with a significant spin down, although the potential of significantly observing such short scale temporal structures is very challenging in reality. 

Due to the absorption by the ejecta from the BNS collision, such flares are to be optimally observed in the energy range from 0.1 to 10 MeV. The estimated detection rate is increasing towards a fluence flux limit, with a power law with an index -1.5, while the rate of such bursts as GW association will also be limited by the detecting reach of GW detector networks, when the fluence limit of the HE detector is below some turn-over sensitivity. Below this turn-over fluence limit, the rate follows another power law with index -2/3. When observing with a detector of energy range 0.1-1\,MeV, the turn-over flux limit is at $\sim2\times10^{-9}$\,erg/cm$^2$, while for a detector of 1-10\,MeV is at $\sim10^{-8}$\,erg/cm$^2$. Based on our evaluation from a population of BNS, a GRB monitor with energy range of $\sim0.1-1$\,MeV and a fluence limit of $\sim2\times10^{-8}$\,erg/cm$^2$ could detect such flares as gamma-ray counterparts of GW events at a rate from 0.01 to 1 per year.
To raise the detection rate of such event to a few to a few tens per year, we expect a future MeV detector working in a range from $\sim1-10$\,MeV with a fluence limit $\sim10^{-9}$\,erg/cm$^2$ for a 1s exposure time.

\begin{acknowledgments}
We would like to acknowledge the insightful discussions we had with Profs. Shuang-Nan Zhang and Ming-Yu Ge. We also appreciate the valuable comments and suggestions from the reviewer. The authors would also like to express  gratitude to Mr. Emre Seyit Yorgancioglu for his proofreading of the manuscript. This work is supported by the National Key R\&D Program of China (2021YFA0718500). SXY acknowledge the support from the Chinese Academy of Sciences (Grant No. E329A3M1). The work of X.W. is supported in part by the Chinese Academy of Sciences (Grant No. E329A6M1). 

\end{acknowledgments}


\begin{thebibliography}{}

\bibitem[Abbott et al.(2017a)]{Abbott} Abbott, B.~P., Abbott, R., Abbott, T.~D., et al.\ 2017, Physical Review Letters, 119, 161101 

\bibitem[Abbott et al.(2017b)]{NSM} Abbott, B.~P., Abbott, R., Abbott, T.~D., et al.\ 2017, \apjl, 848, L12

\bibitem[Abbott et al.(2020)]{2020LRR....23....3A} Abbott, B.~P., Abbott, R., Abbott, T.~D., et al.\ 2020, Living Reviews in Relativity, 23, 3. doi:10.1007/s41114-020-00026-9
\bibitem[Aschwanden(2011)]{2011soca.book.....A} Aschwanden, M.~J.\ 2011, Self-Organized Criticality in Astrophysics, by Markus J. Aschwanden.  Springer-Praxis, Berlin ISBN 978-3-642-15000-5, 416p.

\bibitem[Bak et al.(1987)]{1987PhRvL..59..381B} Bak, P., Tang, C., \& Wiesenfeld, K.\ 1987, \prl, 59, 381. doi:10.1103/PhysRevLett.59.381
\bibitem[Baym \& Pines(1971)]{1971AnPhy..66..816B} Baym, G. \& Pines, D.\ 1971, Annals of Physics, 66, 816. doi:10.1016/0003-4916(71)90084-4
\bibitem[Beloborodov(2020)]{2020ApJ...896..142B} Beloborodov, A.~M.\ 2020, \apj, 896, 142. doi:10.3847/1538-4357/ab83eb 
\bibitem[Blaes et al.(1989)]{1989ApJ...343..839B} Blaes, O., Blandford, R., Goldreich, P., et al.\ 1989, \apj, 343, 839. doi:10.1086/167754 
\bibitem[Blackman \& Yi(1998)]{1998ApJ...498L..31B} Blackman, E.~G. \& Yi, I.\ 1998, \apjl, 498, L31. doi:10.1086/311311
\bibitem[Bovard et al.(2017)]{Bovard} Bovard, L., Martin, D., Guercilena, F., et al.\ 2017, \prd, 96, 124005 

\bibitem[Bransgrove et al.(2020)]{2020ApJ...897..173B} Bransgrove, A., Beloborodov, A.~M., \& Levin, Y.\ 2020, \apj, 897, 173. doi:10.3847/1538-4357/ab93b7 
\bibitem[Burbidge et al.(1957)]{Burbidge1957} Burbidge, E.~M., Burbidge, G.~R., Fowler, W.~A., \& Hoyle, F.\ 1957, Reviews of Modern Physics, 29, 547

\bibitem[Cheng et al.(2020)]{2020MNRAS.491.1498C} Cheng, Y., Zhang, G.~Q., \& Wang, F.~Y.\ 2020, \mnras, 491, 1498. doi:10.1093/mnras/stz3085
\bibitem[Chirenti et al.(2023)]{2023Natur.613..253C} Chirenti, C., Dichiara, S., Lien, A., et al.\ 2023, \nat, 613, 253. doi:10.1038/s41586-022-05497-0
\bibitem[C{\^o}t{\'e} et al.(2018)]{CoteGW170817} C{\^o}t{\'e}, B., Fryer, C.~L., Belczynski, K., et al.\ 2018, \apj, 855, 99. doi:10.3847/1538-4357/aaad67

\bibitem[Cowperthwaite et al.(2017)]{Cowperthwaite2017} Cowperthwaite, P.~S., Berger, E., Villar, V.~A., et al.\ 2017, \apjl, 848, L17 


\bibitem[Dichiara et al.(2013)]{2013ApJ...777..132D} Dichiara, S., Guidorzi, C., Frontera, F., et al.\ 2013, \apj, 777, 132. doi:10.1088/0004-637X/777/2/132
\bibitem[Douchin \& Haensel(2000)]{2000PhLB..485..107D} Douchin, F. \& Haensel, P.\ 2000, Physics Letters B, 485, 107. doi:10.1016/S0370-2693(00)00672-9 
\bibitem[Duncan \& Thompson(1992)]{1992ApJ...392L...9D} Duncan, R.~C. \& Thompson, C.\ 1992, \apjl, 392, L9. doi:10.1086/186413
\bibitem[Fahlman \& Gregory(1981)]{1981Natur.293..202F} Fahlman, G.~G. \& Gregory, P.~C.\ 1981, \nat, 293, 202. doi:10.1038/293202a0

\bibitem[Fern{\'a}ndez et al.(2015)]{Rodrigo2015} Fern{\'a}ndez, R., Kasen, D., Metzger, B.~D., et al.\ 2015, \mnras, 446, 750. doi:10.1093/mnras/stu2112

\bibitem[Ferrario \& Wickramasinghe(2008)]{2008MNRAS.389L..66F} Ferrario, L. \& Wickramasinghe, D.\ 2008, \mnras, 389, L66. doi:10.1111/j.1745-3933.2008.00527.x
\bibitem[Gavriil et al.(2002)]{2002Natur.419..142G} Gavriil, F.~P., Kaspi, V.~M., \& Woods, P.~M.\ 2002, \nat, 419, 142. doi:10.1038/nature01011
\bibitem[Golenetskii et al.(1984)]{1984Natur.307...41G} Golenetskii, S.~V., Ilinskii, V.~N., \& Mazets, E.~P.\ 1984, \nat, 307, 41. doi:10.1038/307041a0
\bibitem[Haensel \& Pichon(1994)]{1994A&A...283..313H} Haensel, P. \& Pichon, B.\ 1994, \aap, 283, 313. doi:10.48550/arXiv.nucl-th/9310003 
\bibitem[Hartmann(1995)]{1995A&ARv...6..225H} Hartmann, D.~H.\ 1995, \aapr, 6, 225. doi:10.1007/BF01837116
\bibitem[Hendriks et al.(2022)]{2022arXiv220814156H} Hendriks, K., Yi, S.-X., \& Nelemans, G.\ 2022, arXiv:2208.14156. doi:10.48550/arXiv.2208.14156
\bibitem[Hurley et al.(2005)]{2005Natur.434.1098H} Hurley, K., Boggs, S.~E., Smith, D.~M., et al.\ 2005, \nat, 434, 1098. doi:10.1038/nature03519
\bibitem[Just et al.(2015)]{Just} Just, O., Bauswein, A., Ardevol Pulpillo, R., et al.\ 2015, \mnras, 448, 541. doi:10.1093/mnras/stv009


\bibitem[Kasen et al.(2017)]{Kasen2017} Kasen, D., Metzger, B., Barnes, J., et al.\ 2017, \nat, 551, 80.


\bibitem[Kaspi \& Gavriil(2004)]{2004NuPhS.132..456K} Kaspi, V.~M. \& Gavriil, F.~P.\ 2004, Nuclear Physics B Proceedings Supplements, 132, 456. doi:10.1016/j.nuclphysbps.2004.04.080
\bibitem[Kaspi \& Beloborodov(2017)]{2017ARA&A..55..261K} Kaspi, V.~M. \& Beloborodov, A.~M.\ 2017, \araa, 55, 261. doi:10.1146/annurev-astro-081915-023329

\bibitem[Kiuchi et al.(2014)]{2014PhRvD..90d1502K} Kiuchi, K., Kyutoku, K., Sekiguchi, Y., et al.\ 2014, \prd, 90, 041502. doi:10.1103/PhysRevD.90.041502

\bibitem[Klu{\'z}niak \& Ruderman(1998)]{1998ApJ...505L.113K} Klu{\'z}niak, W. \& Ruderman, M.\ 1998, \apjl, 505, L113. doi:10.1086/311622
\bibitem[Komissarov et al.(2007)]{2007MNRAS.374..415K} Komissarov, S.~S., Barkov, M., \& Lyutikov, M.\ 2007, \mnras, 374, 415. doi:10.1111/j.1365-2966.2006.11152.x 

\bibitem[Korobkin et al.(2020)]{Oleg} Korobkin, O., Hungerford, A.~M., Fryer, C.~L., et al.\ 2020, \apj, 889, 168. doi:10.3847/1538-4357/ab64d8


\bibitem[Lasky et al.(2011)]{2011ApJ...735L..20L} Lasky, P.~D., Zink, B., Kokkotas, K.~D., et al.\ 2011, \apjl, 735, L20. doi:10.1088/2041-8205/735/1/L20
\bibitem[Lattimer et al.(1994)]{1994ApJ...425..802L} Lattimer, J.~M., van Riper, K.~A., Prakash, M., et al.\ 1994, \apj, 425, 802. doi:10.1086/174025 
\bibitem[Li \& Paczy{\'n}ski(1998)]{1998ApJ...507L..59L} Li, L.-X. \& Paczy{\'n}ski, B.\ 1998, \apjl, 507, L59. doi:10.1086/311680
\bibitem[Levin(2006)]{2006MNRAS.368L..35L} Levin, Y.\ 2006, \mnras, 368, L35.
doi:10.1111/j.1745-3933.2006.00155.x 
\bibitem[Lu \& Hamilton(1991)]{1991ApJ...380L..89L} Lu, E.~T. \& Hamilton, R.~J.\ 1991, \apjl, 380, L89. doi:10.1086/186180
\bibitem[Lyutikov(2003)]{2003MNRAS.346..540L} Lyutikov, M.\ 2003, \mnras, 346, 540. doi:10.1046/j.1365-2966.2003.07110.x 
\bibitem[Mahlmann et al.(2019)]{2019MNRAS.490.4858M} Mahlmann, J.~F., Akg{\"u}n, T., Pons, J.~A., et al.\ 2019, \mnras, 490, 4858. doi:10.1093/mnras/stz2729 

\bibitem[Mereghetti et al.(2015)]{2015SSRv..191..315M} Mereghetti, S., Pons, J.~A., \& Melatos, A.\ 2015, \ssr, 191, 315. doi:10.1007/s11214-015-0146-y
\bibitem[Mereghetti(2008)]{2008A&ARv..15..225M} Mereghetti, S.\ 2008, \aapr, 15, 225. doi:10.1007/s00159-008-0011-z


\bibitem[Metzger \& Berger(2012)]{Metzger} Metzger, B.~D. \& Berger, E.\ 2012, \apj, 746, 48. doi:10.1088/0004-637X/746/1/48

\bibitem[Minaev \& Pozanenko(2020)]{2020AstL...46..573M} Minaev, P.~Y. \& Pozanenko, A.~S.\ 2020, Astronomy Letters, 46, 573. doi:10.1134/S1063773720090042

\bibitem[Mumpower et al.(2018)]{Matt} Mumpower, M.~R., Kawano, T., Sprouse, T.~M., et al.\ 2018, \apj, 869, 14

\bibitem[Nakamura(1998)]{1998PThPh.100..921N} Nakamura, T.\ 1998, Progress of Theoretical Physics, 100, 921. doi:10.1143/PTP.100.921
\bibitem[Narayana Bhat et al.(2016)]{2016ApJS..223...28N} Narayana Bhat, P., Meegan, C.~A., von Kienlin, A., et al.\ 2016, \apjs, 223, 28. doi:10.3847/0067-0049/223/2/28
\bibitem[Negele \& Vautherin(1973)]{1973NuPhA.207..298N} Negele, J.~W. \& Vautherin, D.\ 1973, \nphysa, 207, 298. doi:10.1016/0375-9474(73)90349-7 
\bibitem[Norris et al.(1991)]{1991ApJ...366..240N} Norris, J.~P., Hertz, P., Wood, K.~S., et al.\ 1991, \apj, 366, 240. doi:10.1086/169556
\bibitem[Olami et al.(1992)]{1992PhRvL..68.1244O} Olami, Z., Feder, H.~J.~S., \& Christensen, K.\ 1992, \prl, 68, 1244. doi:10.1103/PhysRevLett.68.1244
\bibitem[Palmer et al.(2005)]{2005Natur.434.1107P} Palmer, D.~M., Barthelmy, S., Gehrels, N., et al.\ 2005, \nat, 434, 1107. doi:10.1038/nature03525
\bibitem[Radice et al.(2018)]{Radice18} Radice, D., Perego, A., Hotokezaka, K., et al., 2018, \apj, 869, 130

\bibitem[Rea et al.(2010)]{2010Sci...330..944R} Rea, N., Esposito, P., Turolla, R., et al.\ 2010, Science, 330, 944. doi:10.1126/science.1196088 
\bibitem[Rea \& Esposito(2011)]{2011ASSP...21..247R} Rea, N. \& Esposito, P.\ 2011, High-Energy Emission from Pulsars and their Systems, 21, 247. doi:10.1007/978-3-642-17251-9\_21
\bibitem[Ripperda et al.(2019)]{2019MNRAS.485..299R} Ripperda, B., Porth, O., Sironi, L., et al.\ 2019, \mnras, 485, 299. doi:10.1093/mnras/stz387 
\bibitem[Roberts et al.(2021)]{2021AAS...23723302R} Roberts, O.~J., Veres, P., Baring, M., et al.\ 2021, \aas
\bibitem[Rosswog et al.(2013)]{rosswog} Rosswog, S., Piran, T., \& Nakar, E.\ 2013, \mnras, 430, 2585. doi:10.1093/mnras/sts708

\bibitem[Rosswog et al.(2018)]{Rosswog2018} Rosswog, S., Sollerman, J., Feindt, U., et al.\ 2018, \aap, 615, A132 

\bibitem[Ruderman et al.(2000)]{2000ApJ...542..243R} Ruderman, M.~A., Tao, L., \& Klu{\'z}niak, W.\ 2000, \apj, 542, 243. doi:10.1086/309537
\bibitem[Thompson(1994)]{1994MNRAS.270..480T} Thompson, C.\ 1994, \mnras, 270, 480. doi:10.1093/mnras/270.3.480
\bibitem[Thompson \& Duncan(1995)]{1995MNRAS.275..255T} Thompson, C. \& Duncan, R.~C.\ 1995, \mnras, 275, 255. doi:10.1093/mnras/275.2.255
\bibitem[Thompson \& Duncan(1996)]{1996ApJ...473..322T} Thompson, C. \& Duncan, R.~C.\ 1996, \apj, 473, 322. doi:10.1086/178147 

\bibitem[Sharma et al. (2023)]{2023arXivS} Sharma, P., Barkov, M., Lyutikov, M. \ 2023,arXiv: 2302.08848

\bibitem[Sneden et al.(2008)]{Sneden2008} Sneden, C., Cowan, J.~J., \& Gallino, R.\ 2008, \araa, 46, 241 

\bibitem[Spruit(1999)]{1999A&A...341L...1S} Spruit, H.~C.\ 1999, \aap, 341, L1. doi:10.48550/arXiv.astro-ph/9811007
\bibitem[Strohmayer et al.(1991)]{1991ApJ...375..679S} Strohmayer, T., Ogata, S., Iyetomi, H., et al.\ 1991, \apj, 375, 679. doi:10.1086/170231 
\bibitem[Svinkin et al.(2021)]{2021Natur.589..211S} Svinkin, D., Frederiks, D., Hurley, K., et al.\ 2021, \nat, 589, 211. doi:10.1038/s41586-020-03076-9

\bibitem[Tanaka \& Hotokezaka(2013)]{Tanaka2013} Tanaka, M., \& Hotokezaka, K.\ 2013, \apj, 775, 113 

\bibitem[The LIGO Scientific Collaboration et al.(2021)]{2021arXiv211103634T} The LIGO Scientific Collaboration, the Virgo Collaboration, the KAGRA Collaboration, et al.\ 2021, arXiv:2111.03634. doi:10.48550/arXiv.2111.03634
\bibitem[Turolla et al.(2015)]{2015RPPh...78k6901T} Turolla, R., Zane, S., \& Watts, A.~L.\ 2015, Reports on Progress in Physics, 78, 116901. doi:10.1088/0034-4885/78/11/116901
\bibitem[Turolla \& Esposito(2013)]{2013IJMPD..2230024T} Turolla, R. \& Esposito, P.\ 2013, International Journal of Modern Physics D, 22, 1330024-163. doi:10.1142/S0218271813300243 
\bibitem[Usov(1992)]{1992Natur.357..472U} Usov, V.~V.\ 1992, \nat, 357, 472. doi:10.1038/357472a0

\bibitem[Wang et al.(2020a)]{Wang2020a} Wang, X., N3AS Collaboration, Fields, B. D., et al.\ 2020, \apj, 893, 92. doi:10.3847/1538-4357/ab7ffd

\bibitem[Wang et al.(2020b)]{Wang2020b} Wang, X., N3AS Collaboration, Vassh, N., et al.\ 2020, \apjl, 903, L3. doi:10.3847/2041-8213/abbe18

\bibitem[Watson et al.(2019)]{Watson2019} Watson, D., Hansen, C.~J., Selsing, J., et al.\ 2019, \nat, 574, 497

\bibitem[Wheeler et al.(2000)]{2000ApJ...537..810W} Wheeler, J.~C., Yi, I., H{\"o}flich, P., et al.\ 2000, \apj, 537, 810. doi:10.1086/309055

\bibitem[Wollaeger et al.(2018)]{Wollaeger2018} Wollaeger, R.~T., Korobkin, O., Fontes, C.~J., et al.\ 2018, \mnras, 478, 3298

\bibitem[Yi \& Blackman(1998)]{1998ApJ...494L.163Y} Yi, I. \& Blackman, E.~G.\ 1998, \apjl, 494, L163. doi:10.1086/311192
\bibitem[Yuan et al.(2020)]{2020ApJ...900L..21Y} Yuan, Y., Beloborodov, A.~M., Chen, A.~Y., et al.\ 2020, \apjl, 900, L21. doi:10.3847/2041-8213/abafa8 
\bibitem[Zhang \& M{\'e}sz{\'a}ros(2001)]{2001ApJ...552L..35Z} Zhang, B. \& M{\'e}sz{\'a}ros, P.\ 2001, \apjl, 552, L35. doi:10.1086/320255
\bibitem[Zhang et al.(2020)]{2020ApJ...903L..32Z} Zhang, H.-M., Liu, R.-Y., Zhong, S.-Q., et al.\ 2020, \apjl, 903, L32. doi:10.3847/2041-8213/abc2c9

\bibitem[Zhang et al.(2022)]{ZYetal2022} Zhang, Z., Yi, S.-X., Zhang, S.-N., Xiong, S.-L., and Xiao, S., 2022, \apjl, 939, L25. doi:10.3847/2041-8213/ac9b55












%

%
%
%




\end{thebibliography}
\end{document}